# The probability of improvement in Fisher's geometric model: a probabilistic approach

## Yoav Ram and Lilach Hadany

The Department of Molecular Biology and Ecology of Plants

The George S. Wise Faculty of Life Sciences

Tel-Aviv University, Tel-Aviv 69978, Israel

Corresponding author: Yoav Ram, yoavram@post.tau.ac.il, +972.3.640.6886

# Abstract

Fisher developed his geometric model to support the *micro-mutationalism* hypothesis which claims that small mutations are more likely to be beneficial and therefore to contribute to evolution and adaptation. While others have provided a general solution to the model using geometric approaches, we derive an equivalent general solution using a probabilistic approach. Our approach to Fisher's geometric model provides alternative intuition and interpretation of the solution in terms of the model's parameters: for mutation to improve a phenotype, its relative beneficial effect must be larger than the ratio of its total effect and twice the difference between the current phenotype and the optimal one. Our approach provides new insight into this classical model of adaptive evolution.

# 1 Introduction

Fisher's geometric model (FGM) is a widely used model of adaptive evolution in which selection and mutation act on a combination of quantitative traits. Each trait has an optimal value, and the fitness of trait combinations is a decreasing function of the distance to the optimal trait combination. The model was originally used by Fisher to calculate the probability that a pleiotropic mutation - one that affects multiple traits - leads to an improved phenotype. In support of the *micro-mutationalism* hypothesis, Fisher found that small mutations are more likely to be beneficial and therefore contribute to adaptation and evolution (Fisher, 1930, p. 40; Waxman and Welch, 2005).

FGM is very relevant to both theoretical and experimental research in evolutionary biology. FGM has been used to infer distributions of fitness effects and fitness landscapes (Blanquart et al., 2014; Hietpas et al., 2013; MacLean et al., 2010; Melnyk and Kassen, 2011; Orr, 1998; Sousa et al., 2012; Trindade et al., 2012; Weinreich and Knies, 2013). Perfeito et al. (2014) followed the adaptation of 23 lines of *E. coli* and used Approximate Bayesian Computation to fit the data to the parameters of FGM, including the genomic mutation rate, number of traits, and the mean phenotypic effect of mutations. Bank et al. (Bank et al., 2014) estimated the distribution of fitness effects of 560 point mutations in *Hsp90* in *Saccharomyces cerevisiae* in six environments. Their results agree with predictions of the FGM. FGM was also used to test the *micro-mutationalism* hypothesis (Burch and Chao, 1999) and for analyzing evolutionary dynamics with simulations (Matuszewski et al., 2014; Venkataram et al., 2013). It has been extended to include fixation probabilities, fitness functions, and epistasis (Martin and Lenormand, 2008, 2006; Waxman, 2006). Finally, a recent article by Martin (2014) provides a biological justification for FGM by demonstrating its emergence in complex phenotypic networks.

Previous derivations of the probability of improvement in FGM used geometric approaches (Hartl and Taubes, 1996; Rice, 1990). Here, we study this problem using a probabilistic approach. Our result provides an alternative interpretation of how the probability of improvement in the phenotype after occurrence of a mutation is affected by the model's parameters: the total effect of mutation on phenotype, the number of affected traits, and the difference between the current phenotype and the optimal one. Additionally, we demonstrate how our approach can be used to analyze other properties of adaptation in FGM.

# 2 Model

We start by describing the general FGM. Then we review previously published results of the probability for improvement with two and three traits, with an arbitrary number of traits, and with a large number of traits.

## 2.1 Overview

In the following, we use the notation introduced by Fisher (1930, p. 40). In FGM, a phenotype is defined by *n* traits and therefore can be described by a vector in an *n*-dimensional space ($\mathbb{R}^n$). Because we are interested in the effect of mutation on phenotype, we define the Euclidean distance of the current phenotype from the optimal phenotype to be *d/2*. Without loss of generality, we assume that the optimal trait combination is *O=(0, …, 0)* and the current phenotype is *A=(d/2, 0, …, 0)*. The effect of mutation on the current phenotype is given by a vector of magnitude *r* and a random direction.

Fisher's goal was to calculate the probability *p* that a mutation is beneficial - that is, that a mutation creates a mutant phenotype which is closer to the optimal phenotype then the current one.

## 2.2 Review of previous results

### 2.2.1 Two traits

Figure 1 illustrates the model for two traits. We denote the current phenotype by *A=(d/2,0)*. Phenotypes that can be reached by a single mutation of size *r* lie on circle α (centered at *A* with radius *r*). Phenotypes that are as fit as *A* are marked by circle β (centered at the origin with radius *d/2*). For a mutation to be beneficial, the mutant phenotype must be in the part of the circle α that is inside circle β (the dashed arc).

The two circles intersect at *B=(x,y)* and *(x,-y)* and we define *C=(x,0)*. *x* is calculated using the two circle equations:

$$\begin{cases} \left(x - \dfrac{d}{2}\right)^2 + y^2 = r^2 \\ x^2 + y^2 = \left(\dfrac{d}{2}\right)^2 \end{cases} \Rightarrow$$

$$\begin{cases} x = \dfrac{d}{2} - \dfrac{r^2}{d} \\ y = \pm\sqrt{r^2 - \dfrac{r^4}{d^2}} \end{cases}.$$

Therefore, the length of AC is $r^2/d$. AB is the radius of circle α with length $r$, and θ is the angle between CA and AB. Therefore, $\cos\theta = AC/AB = r/d$. The probability of improvement $p$ is the ratio between the (dashed) arc of circle α that lies inside circle β (with an angle 2θ) and the whole circle (with an angle 2π). For this two-dimensional case, the final formula for the beneficial mutation probability is (Rice, 1990)

$$p_2 = \frac{1}{\pi}\cos^{-1}\left(\frac{r}{d}\right),$$

Where $cos^{-1}x$ is arccos, the inverse function of cosine.

### 2.2.2 Three traits

With three traits, we have two spheres: sphere β, centered at O=(0,0,0) with radius $d/2$, and sphere α centered at *(d/2,0,0)* with radius *r*.

The intersection of these spheres defines a plane that cuts the sphere α to create a spherical cap. The area of this spherical cap has a simple formula, $2\pi rh$, where $h$ is the height of the cap (equivalent to *r-AC* in Figure 1). This height can be found by the same way as in the two trait case: *h=r(1-r/d)*.

Because the entire area of the surface of sphere α is $4\pi r^2$, the ratio between the area of the spherical cap and the whole sphere is (Fisher, 1930, p. 40)

$$p_3 = \frac{2\pi r^2\left(1-\frac{r}{d}\right)}{4\pi r^2} = \frac{1}{2}\left(1-\frac{r}{d}\right).$$

### 2.2.3 Arbitrary number of traits

In the general case of *n* traits there are two *n-1* hyperspheres, but the rest of the details are similar to the n=2,3 cases. Rice (1990) presented a solution based on an argument similar to that made in the three trait case on spherical caps:

$$p_n^R = \frac{\int_0^{\cos^{-1}\left(\frac{r}{d}\right)} \sin^{n-2}(\theta)d\theta}{\int_0^{\pi} \sin^{n-2}(\theta)d\theta}.$$

This result was also derived by others (Hartl and Taubes, 1996; Waxman and Welch, 2005). For *n*=2 this becomes

$$\frac{\int_0^{\cos^{-1}\left(\frac{r}{d}\right)} 1\, d\theta}{\int_0^{\pi} 1\, d\theta} = \frac{1}{\pi}\cos^{-1}\left(\frac{r}{d}\right) = p_2.$$

For *n*=3, this becomes

$$\frac{\int_0^{\cos^{-1}\left(\frac{r}{d}\right)} \sin(\theta)\, d\theta}{\int_0^{\pi} \sin(\theta)\, d\theta} = \frac{-\cos\left(\cos^{-1}\left(\frac{r}{d}\right)\right) + \cos(0)}{-\cos(\pi) + \cos(0)} = \frac{1}{2}\left(1 - \frac{r}{d}\right) = p_3.$$

Kimura (1983, p. 137) presented a different formula (which is equivalent, see below):

$$p_n^K = \frac{1}{2} I_{1-\frac{r^2}{d^2}}\left(\frac{n-1}{2}, \frac{1}{2}\right),$$

where $I_x(a, b)$ is the regularized incomplete beta function. He did not provide a derivation for this result.

2.2.4    Large number of traits

Fisher presented an asymptotic result for a large number of traits (Fisher, 1930, p. 40), which is very elegant due to the use of $\phi$, the cumulative distribution function of the standard normal distribution:

$$p_\infty = 1 - \phi\left(\sqrt{n}\frac{r}{d}\right) = \frac{1}{\sqrt{2\pi}} \int_{\sqrt{n}\frac{r}{d}}^{\infty} \exp\left(-\frac{t^2}{2}\right) dt.$$

Fisher did not include a derivation of the result, but it can be calculated as an approximation of the general solution $p_n^R$ (Hartl and Taubes, 1996; Waxman and Welch, 2005). This is a good approximation even for an intermediate number of traits such as *n*=9 (Kimura, 1983, fig. 6.5).

# 3   Results

The formulas in the Model section were derived using analytic geometry. However, the general ($p_n^R, p_n^K$) and asymptotic ($p_\infty$) equations are not easily interpreted in relation to the model's parameters.

Here, we use a probabilistic approach to derive the probability that a mutation is beneficial and demonstrate that our result is equivalent to previously published results. The advantage of the probabilistic approach is that the result can be more easily understood in terms of the model's parameters. Thus, our approach offers an intuitive interpretation of the probability of improvement in FGM. Moreover, we use our probabilistic approach to evaluate the expected phenotypic change towards the optimum and the expected fitness improvement in beneficial mutations, demonstrating that our approach can be used to analyze additional properties of the model.

## 3.1  The probabilistic approach

There are several approaches for choosing random points on a hypersphere with radius $r$ in an $n$-dimensional space so that the points will be uniformly distributed. The simplest, perhaps, uses this procedure (Muller, 1959):

1. Generate a random vector $Z = (Z_1, \ldots, Z_n)$ by drawing $n$ independent samples from a standard normal distribution: $Z_i \sim N(0,1), \ 1 \leq i \leq n$.
2. Normalize this vector and multiply it by the desired radius: $X_i = r \cdot Z_i / \|Z\|$.
3. The resulting vector $X = (X_1, \ldots, X_n)$ is a uniformly random point on the surface of a hypersphere with radius $r$.

Note that $\|\cdot\|$ is the Euclidean norm: $\|X\| = \sqrt{X_1^2 + \cdots X_n^2}$.

Given a phenotype *A* and a mutation *X*, what is the probability *p* that the mutant phenotype *A+X* is closer to the optimal phenotype *O* than the original phenotype *A* was?

In probabilistic terms, we are looking for the probability that the Euclidean norm of *A+X* is smaller than that of *A* (which is equal to *(d/2)²*):

$$p = P(\|A + X\| < \|A\|) = P(\|A + X\|^2 < \|A\|^2) = P\left(\sum_{i=1}^{n}(A + X)_i^2 < \left(\frac{d}{2}\right)^2\right).$$

Now, *(A+X)ᵢ=Aᵢ+Xᵢ*, and *A=(d/2, 0, …, 0)*,

$$p = P\left(\left(\frac{d}{2} + X_1\right)^2 + \sum_{i=2}^{n} X_i^2 < \left(\frac{d}{2}\right)^2\right) = P\left(\frac{d^2}{4} + dX_1 + X_1^2 + \sum_{i=2}^{n} X_i^2 < \frac{d^2}{4}\right) =$$

$$P\left(dX_1 + \sum_{i=1}^{n} X_i^2 < 0\right),$$

and by definition $\sum_{i=1}^{n} X_i^2 = \|X\|^2 = r^2$,

$$p = P(X_1 d + r^2 < 0) = P\left(X_1 < -\frac{r^2}{d}\right) = P\left(\frac{X_1}{r} < -\frac{r}{d}\right) =$$

$$P\left(\frac{X_1}{r} > \frac{r}{d}\right), \qquad (1)$$

because $X_1 = r\, Z_1/\|Z\|$ is symmetric around zero due to the symmetry of the standard normal distribution. In words, we found that the probability that a mutation improves the phenotype is the probability that the relative mutation size towards the optimum is greater than the relation between the total size of the mutation and twice the difference between the phenotype and the optimum.

**3.2   Relation to previous results**

To relate this result to the results described in the Model section, we note that eq. (1) can be written as

$$p = P\left(\frac{X_1}{r} > \frac{r}{d}\right) = P\left(\frac{Z_1}{\|Z\|} > \frac{r}{d}\right)$$

by the definition of $X_1$. Now, because $r/d>0$,

$$p = P\left(Z_1 > 0 \text{ and } \frac{Z_1^2}{\|Z\|^2} > \frac{r^2}{d^2}\right) =$$

$$P(Z_1 > 0) \cdot P\left(\frac{Z_1^2}{\|Z\|^2} > \frac{r^2}{d^2}\right), \qquad (2)$$

because $Z_1^2$ is independent of the sign of $Z_1$ due to the symmetry of the standard normal distribution. Now because $P(Z_1 > 0) = \frac{1}{2}$,

$$p = P\left(\frac{X_1}{r} > \frac{r}{d}\right) = \frac{1}{2} P\left(\frac{Z_1^2}{\|Z\|^2} > \frac{r^2}{d^2}\right). \qquad (3)$$

The sum of *k* squares of independent standard normal random variables is a chi-squared random variable with parameter *k*, so $Z_1^2 \sim \chi^2(1)$ and $\|Z\|^2 - Z_1^2 = \sum_{i=2}^{n} Z_i^2 \sim \chi^2(n-1)$. The ratio of one chi-squared to its sum with another independent chi-squared is beta distributed (Dutka, 1981), so we find that $\frac{Z_1^2}{\|Z\|^2} \sim Beta\left(\frac{1}{2}, \frac{n-1}{2}\right)$.

Now, we use the cumulative distribution function of a beta distribution to get the formula given by Kimura (1983, p. 137):

$$p = \frac{1}{2} P\left(\frac{Z_1^2}{\|Z\|^2} > \frac{r^2}{d^2}\right) = \frac{1}{2}\left(1 - P\left(\frac{Z_1^2}{\|Z\|^2} < \frac{r^2}{d^2}\right)\right) =$$

$$\frac{1}{2}\left(1 - I_{\frac{r^2}{d^2}}\left(\frac{1}{2}, \frac{n-1}{2}\right)\right) = \frac{1}{2} I_{1-\frac{r^2}{d^2}}\left(\frac{n-1}{2}, \frac{1}{2}\right) = p_n^K, \quad (4)$$

where $I_x(a, b)$ is the regularized incomplete beta function and the final equality is due to the identity $I_x(a, b) = 1 - I_{1-x}(b, a)$ (Dutka, 1981).

Gale (1990, p. 188) interpreted $p_n^K$ (eq. (4)) as the probability of exceeding $T = \sqrt{(n-1)\frac{r^2/d^2}{1-r^2/d^2}}$ in a Student's t-distributed random variable $Y$ with $n$-1 degrees of freedom. When $n$=2, $Y$ is Cauchy distributed random variable, and

$$P(Y > T) = \frac{1}{2} - \frac{1}{\pi} \tan^{-1}\left(\sqrt{\frac{r^2/d^2}{1-r^2/d^2}}\right) =$$

$$\frac{1}{2} - \frac{1}{\pi} \sin^{-1} \frac{r}{d} = \frac{1}{\pi} \cos^{-1} \frac{r}{d} = p_2,$$

using the identities $\tan^{-1} x = \sin^{-1} \frac{x}{\sqrt{1+x^2}}$ and $\sin^{-1} x = \frac{\pi}{2} - \cos^{-1} x$.

When $n$=3, the cumulative distribution function of $Y$ is $F(x) = \frac{1}{2}\left(1 + \frac{x}{\sqrt{2+x^2}}\right)$ and

$$P(Y > T) = \frac{1}{2}\left(1 - \sqrt{2 \frac{r^2/d^2}{1-r^2/d^2}} \bigg/ \sqrt{2 + 2\frac{r^2/d^2}{1-r^2/d^2}}\right) =$$

$$\frac{1}{2}\left(1 - \frac{r}{d}\right) = p_3.$$

When $n$ is very large, $Y$ can be approximated with a normal random variable:

$$P(Y > T) \approx 1 - \phi\left(\sqrt{(n-1)\frac{r^2/d^2}{1-r^2/d^2}}\right) \approx 1 - \phi\left(\sqrt{n}\frac{r}{d}\right) = p_\infty,$$

where $n - 1 \approx n$ and $d^2 \gg r^2$ for $p$ that isn't too small (Gale, 1990, p. 188).

We will now arrive to eq. (4) from the other general solution $p_n^R$ (see Model section). Following (Li, 2011) we use the identity

$$\int_0^\phi \sin^n(\theta)\, d\theta = \frac{1}{2} B_{\sin^2(\phi)}\left(\frac{n+1}{2}, \frac{1}{2}\right) = \frac{1}{2} I_{\sin^2(\phi)}\left(\frac{n+1}{2}, \frac{1}{2}\right) B\left(\frac{n+1}{2}, \frac{1}{2}\right), \quad (5)$$

where $B_{x(a,b)} \equiv \int_0^x t^{a-1}(1-t)^{b-1} dt$ is the incomplete beta function, related to its regularized form by $I_x(a,b) \equiv \frac{B_x(a,b)}{B(a,b)}$, and $B(a,b) = B_1(a,b)$ is the complete beta function (Dutka, 1981).

We substitute the identity from eq. (5) into the general solution $p_n^R$ to get

$$p_n^R = \frac{\int_0^{\cos^{-1}(\frac{r}{d})} \sin^{n-2}(\theta)\, d\theta}{2\int_0^{\frac{\pi}{2}} \sin^{n-2}(\theta)\, d\theta} = \frac{\frac{1}{2} I_{1-\frac{r^2}{d^2}}\left(\frac{n-1}{2}, \frac{1}{2}\right) B\left(\frac{n-1}{2}, \frac{1}{2}\right)}{I_1\left(\frac{n-1}{2}, \frac{1}{2}\right) B\left(\frac{n-1}{2}, \frac{1}{2}\right)} =$$

$$\frac{1}{2} I_{1-\frac{r^2}{d^2}}\left(\frac{n-1}{2}, \frac{1}{2}\right) = p_n^K,$$

which is the expression in eq. (4).

### 3.3 Asymptotic result

The asymptotic result $p_\infty$, given by Fisher (1930, p. 40) and described in the Model section, can also be derived using the probabilistic approach.

We use the identity (Olver et al., 2010)

$$I_x(a,a) = \frac{1}{2} I_{4x(1-x)}\left(a, \frac{1}{2}\right), \qquad 0 \leq x \leq \frac{1}{2},$$

in eq. (4) by substituting $a = \frac{n-1}{2}$ and solving $4x(1-x) = 1 - \frac{r^2}{d^2} \Rightarrow x = \frac{1}{2}\left(1 - \frac{r}{d}\right)$ (the other solution for $x$ is not relevant as we require $4x(1-x) < 1$):

$$p_n^K = \frac{1}{2} I_{1-\frac{r^2}{d^2}}\left(\frac{n-1}{2}, \frac{1}{2}\right) = I_{\frac{1}{2}(1-\frac{r}{d})}\left(\frac{n-1}{2}, \frac{n-1}{2}\right) = P\left(D < \frac{1}{2}\left(1 - \frac{r}{d}\right)\right),$$

where $D \sim Beta\left(\frac{n-1}{2}, \frac{n-1}{2}\right)$ is a new beta random variable. Beta distributions with equal large parameters can be approximated using a normal distribution with the same expectation and variance (Alfers and Dinges, 1984). We calculate the expectation ($E$) and variance ($V$) according to the appropriate formulas for a beta distribution:

$$E[D] = \frac{1}{2}$$
$$V[D] = \frac{1}{4n}.$$

The normal approximation is therefore given by the normal distribution $N\left(\frac{1}{2}, \frac{1}{4n}\right)$, and we get the asymptotic result:

$$P\left(D < \frac{1}{2}\left(1-\frac{r}{d}\right)\right) \xrightarrow[n\to\infty]{} \phi\left(\frac{\frac{1}{2}\left(1-\frac{r}{d}\right)-E[D]}{\sqrt{V[D]}}\right) = \phi\left(-\sqrt{n}\frac{r}{d}\right) = 1 - \phi\left(\sqrt{n}\frac{r}{d}\right) = p_\infty.$$

As a side note, when *n*=2, $D \sim Beta\left(\frac{1}{2}, \frac{1}{2}\right)$ which is equivalent to the arcsine distribution, and when *n*=3, $D \sim Beta(1,1)$ which is equivalent to a uniform distribution on *(0,1)*. These equivalencies easily return the expressions for $p_2$ and $p_3$.

### 3.4    Expected improvement and fixation probability

The *micro-mutationalism* hypothesis suggests that evolution proceeds through the accumulation of small, relatively frequent mutations, rather than large, rare mutations. Fisher's result, $p_\infty = 1 - \phi(\sqrt{n}\frac{r}{d})$, demonstrated that $p_\infty$ the probability that a mutation improves the phenotype decreases as the mutation size *r* increases. This is because $\phi$ is a cumulative distribution function and is therefore increasing in *r*. This result supported the *micro-mutationalism* hypothesis (Fisher, 1930, p. 40; Waxman and Welch, 2005). However, Kimura (1983, pp. 154–155; Orr, 1998) claimed that considering the probability that a mutation is beneficial isn't sufficient; one must also consider the probability that a beneficial mutation avoids extinction by genetic drift (Patwa and Wahl, 2008). The fixation probability of a beneficial mutation in a large population can be approximated by 2*s*, twice the selective advantage of the beneficial mutation (Eshel, 1981). Therefore the probability that a mutation improves fitness and avoids extinction is:

$$p_{fix} \approx 2sp. \tag{6}$$

In the FGM scenario, the selective advantage of the mutant phenotype is (Figure 1):

$$s = \frac{\omega(A+X)}{\omega(A)} - 1, \tag{7}$$

where $\omega(Z)$ is the fitness of phenotype $Z$. Therefore, evaluation of the fixation probability requires an additional assumption about the fitness function $\omega$. We will begin by exploring the expected phenotypic change towards the optimum in beneficial mutations, as this does not depend on a choice of a fitness function:

$$E[|X_1|||A+X\| < \|A\|] = E\left[|X_1|\left|\frac{X_1}{r} > \frac{r}{d}\right.\right] = rE\left[\frac{|Z_1|}{\|Z\|}\left|\frac{Z_1}{\|Z\|} > \frac{r}{d}\right.\right]$$

$$= rE\left[\sqrt{\frac{Z_1^2}{\|Z\|^2}}\left|\frac{Z_1^2}{\|Z\|^2} > \frac{r^2}{d^2} \text{ and } Z_1 > 0\right.\right].$$

As before, $\frac{Z_1^2}{\|Z\|^2} \sim Beta\left(\frac{1}{2}, \frac{n-1}{2}\right)$ and the sign of $Z_1$ is independent of $\frac{Z_1^2}{\|Z\|^2}$. Therefore, denoting the probability density function of the beta distribution by $f$, and accounting for the constraint that $Z_1 > 0$:

$$E[|X_1|||A+X\| < \|A\|] = r \cdot \frac{1}{2}\int_{\frac{r^2}{d^2}}^{1} \sqrt{z}\, f(z)\, dz \cdot p_n^{-1} =$$

$$r \cdot \frac{1}{2}\int_{\frac{r^2}{d^2}}^{1} z^{\frac{1}{2}} z^{-\frac{1}{2}} (1-z)^{\frac{n-3}{2}} dz \cdot B\left(\frac{1}{2}, \frac{n-1}{2}\right)^{-1} \cdot 2I_{1-\frac{r^2}{d^2}}\left(\frac{n-1}{2}, \frac{1}{2}\right)^{-1}.$$

The integrand becomes $(1-z)^{\frac{n-3}{2}}$ which has an anti-derivative $\frac{-2}{n-1}(1-z)^{\frac{n-1}{2}}$. The two beta functions can be manipulated, using the symmetry of $B(a,b)$ and the definition of $I_x(a,b)$ so that

$$E[|X_1|||A+X\| < \|A\|] = \frac{-2r}{n-1}\left[(1-z)^{\frac{n-1}{2}}\right]_{\frac{r^2}{d^2}}^{1} \cdot B_{1-\frac{r^2}{d^2}}\left(\frac{n-1}{2}, \frac{1}{2}\right)^{-1},$$

and therefore

$$E[|X_1|||A+X\| < \|A\|] = \frac{2r}{(n-1)B_{1-\frac{r^2}{d^2}}\left(\frac{n-1}{2}, \frac{1}{2}\right)}\left(1-\frac{r^2}{d^2}\right)^{\frac{n-1}{2}}. \tag{8}$$

Next, the expected selective advantage $s$ of beneficial mutations, or the improvement in fitness, is, using eq. (7):

$$E[s|s>0] = E\left[\frac{\omega(A+X)}{\omega(A)} - 1\middle|\omega(A+X) > \omega(A)\right].$$

To proceed we must choose a fitness function. A common one in the literature is (Martin and Lenormand, 2006; Orr, 1998)

$$\omega(V) = \exp(-\|V\|^2) \approx 1 - \|V\|^2, \tag{9}$$

where the approximation is valid for phenotypes close to the optimum ($\|V\|^2 \ll 1$). We will use this approximation because it allows us to reach a closed form expression. So,

$$E[s|s > 0] = E\left[\frac{1 - \|A + X\|^2}{1 - \|A\|^2} - 1 \middle| 1 - \|A + X\|^2 > 1 - \|A\|^2\right],$$

and using the definition of the norm we find:

$$E[s|s > 0] = E\left[\frac{1 - \frac{d^2}{4} - dX_1 - r^2}{1 - \frac{d^2}{4}} - 1 \middle| \|A + X\| < \|A\|\right] \Rightarrow$$

$$= E\left[\frac{-dX_1 - r^2}{1 - \frac{d^2}{4}} \middle| \|A + X\| < \|A\|\right].$$

Because $s>0$ requires $X_1<0$ (the mutation must decrease the distance to the optimum), we substitute $-E[X_1|\|A + X\| < \|A\|] = E[|X_1|\|A + X\| < \|A\|]$ to get:

$$E[s|s > 0] = \frac{dE[|X_1|\|A + X\| < \|A\|] - r^2}{1 - \frac{d^2}{4}}, \tag{10}$$

where $E[|X_1|\|A + X\| < \|A\|]$ can be evaluated using eq. (8). Now the fixation probability can be approximated by substituting eqs. (4) and (10) in eq. (6):

$$p_{fix} \approx 2 \cdot E[s|s > 0] \cdot p_n^K.$$

Figure 2 shows the fixation probability $p_{fix}$ as a function of the ratio between the mutation size and the difference from the optimum, $r/d$, for $n$=10. The optimal mutation size is $r \approx 0.2d$. Smaller mutations are more likely to drift to extinction, while larger mutations are less likely to improve the phenotype. Therefore, we can expect mutations of intermediate size to dominate the adaptation process (Kimura, 1983, p. 155).

## 4 Discussion

We have obtained a new formula for the probability of improvement in Fisher's geometric model (eq. (1)): for a mutation with effect *r* on a phenotype at a distance *d/2* from the optimum, the probability that the mutation is beneficial is the probability that $\frac{X_1}{r} > \frac{r}{d}$, where $X_1$ is the effect of the mutation on the phenotype in the direction towards the optimum (Figure 3).

Our result is valuable because it is easily explained in terms of the model's parameters: for a mutation to improve fitness, its relative beneficial effect on the phenotype ($X_1/r$) must be larger than the ratio of its total effect on the phenotype (*r*) and twice the difference from the optimal phenotype (*d*). See Figure 3 for an illustration.

From this description, it is clear that the probability of improvement increases with the difference from the optimal phenotype and decreases with the total mutation effect on the phenotype. Moreover, when the total mutation effect is very small related to the difference from the optimum, almost any change towards the optimum is enough for the mutation to improve the phenotype as long as the mutation effect is in the right direction – which happens in half of the mutations. What about the number of traits *n*? Recall that $\|X\| = \sqrt{\sum_{i=1}^{n} X_i^2} = r$, so that the effect of the mutation on phenotype *r* has to be distributed among all *n* traits. Therefore, for a fixed mutation effect *r*, higher *n* entails a smaller chance that the fraction of the mutation that affects the phenotype in the direction of the optimum $X_1/r$ is large enough to exceed the threshold *r/d*.

We have also used our probabilistic approach to evaluate the expected change towards the optimum in beneficial mutations and, given a fitness function, the expected fixation probability of a beneficial mutation (Figure 2). Our results are consistent with the literature and demonstrate that the probabilistic approach is useful for analysis of other properties of FGM.

In conclusion, Fisher's geometric model is an important and intuitive tool that facilitates the understanding of evolution and adaptation. Our probabilistic approach supplements the literature on FGM by providing a different way to understand, interpret, and work with this important model.

# 5   Acknowledgments

We thank D. Gilat, A. Ryvkin, M. Broido, I. Ben-Zion, U. Obolski, N. Rosenberg, and two anonymous reviewers for helpful discussions and/or comments on the manuscript. This research has been supported in part by the Israeli Science Foundation 1568/13 (LH) and by a fellowship from the Manna Program in Food Safety and Security (YR).

# 7 Figure captions

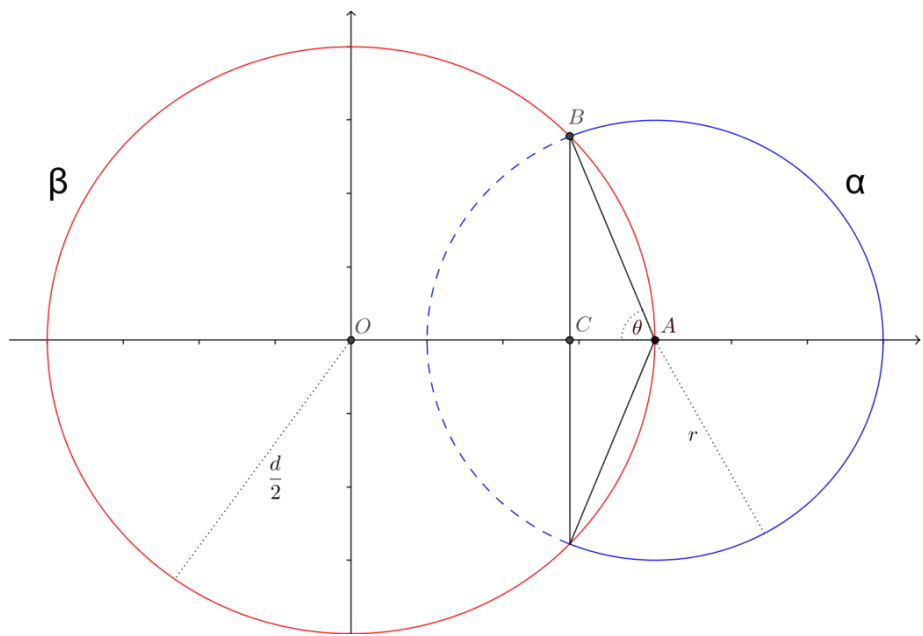

**Figure 1 – Two dimensional presentation of Fisher's geometric model.** The probability of improvement is the probability that a mutation changes the current phenotype *A* to a fitter phenotype that is closer to the optimum *O*. This probability is equal to the fraction of circle α that is inside the circle β (the dashed arc), because the circle α contains all the phenotypes that can be reached by a single mutation in *A* and the circle β contains all the phenotypes that are as fit as *A*.

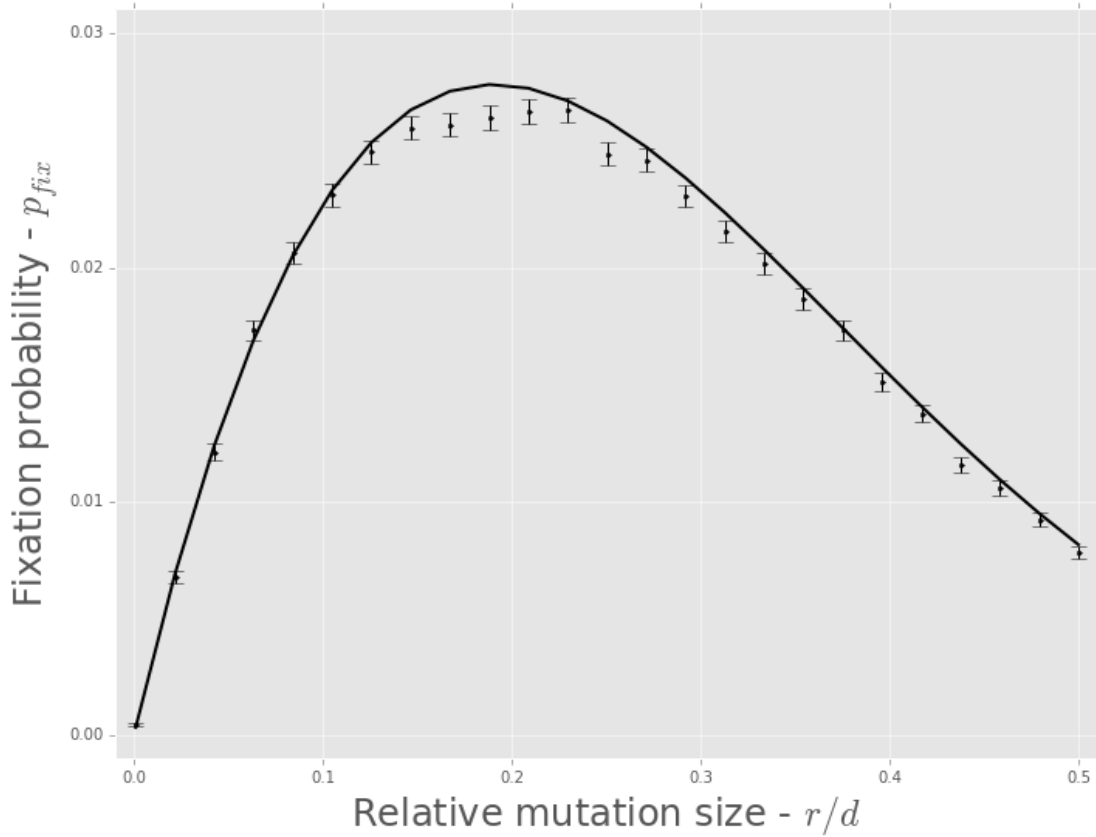

**Figure 2 – Probability of fixation of a beneficial mutation.** The probability that a mutation improves the phenotype and survives extinction by genetic drift, as a function of the ratio between the mutation size and twice the difference from the optimum. The line represents eq. (6) with eqs. (4) and (10) substituted for *p* and *s*, respectively. The markers represent results of stochastic simulations (see Supporting Material). Error bars represent the standard error of the mean. Compare with (Kimura, 1983, fig. 7.2; Orr, 1998, fig. 2b).

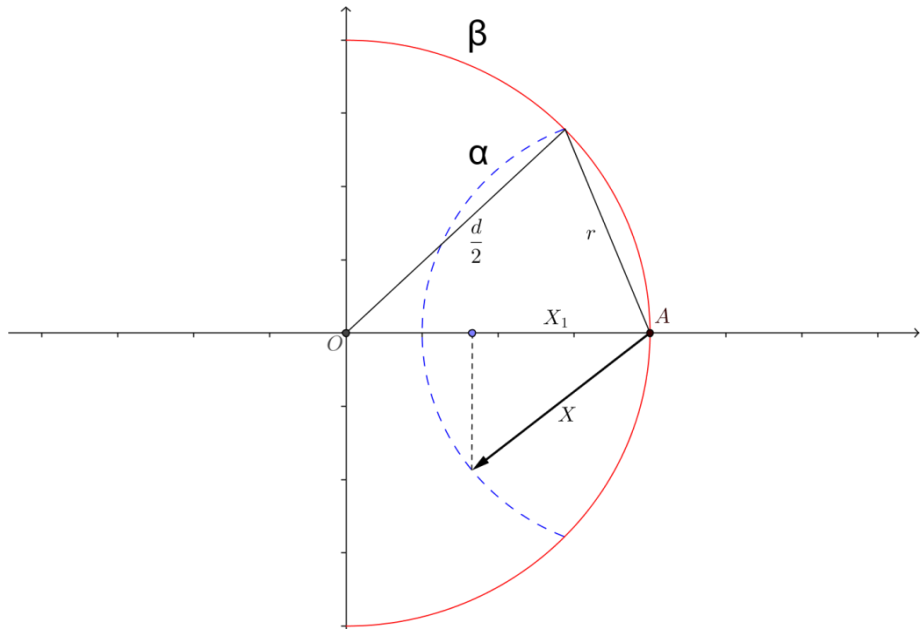

**Figure 3 –Illustration of the probability of improvement.** The probability of improvement can be calculated by drawing a random point on circle α and checking if it is within the circle β. The beneficial effect of a mutation *X* on the phenotype *A* is given by $X_1$. The relative effect of the mutation devoted to the beneficial effect ($X_1/r$) must be larger than the ratio between the effect of the mutation and twice the difference between the current and the optimal phenotypes (*r/d*).